\documentclass[twocolumn,twoside,superscriptaddress,pra,aps,showpacs]{revtex4}
\usepackage{}
\usepackage{amsmath,mathrsfs,amsbsy,color,graphicx,bm,amsthm,amsfonts}
\usepackage{units}
\usepackage{bbm}
\usepackage{times}
\usepackage{epsfig,graphicx}
\usepackage{dcolumn}

\usepackage[latin2]{inputenc}

\begin{document}

\title{The general fine-grained uncertainty relation and the second law of thermodynamics}
\author{Li-Hang Ren}
\affiliation{Beijing National Laboratory for Condensed Matter Physics, Institute of
Physics, Chinese Academy of Sciences, Beijing 100190, China}
\author{Heng Fan}
\email{hfan@iphy.ac.cn}
\affiliation{Beijing National Laboratory for Condensed Matter Physics, Institute of
Physics, Chinese Academy of Sciences, Beijing 100190, China}
\affiliation{Collaborative Innovation Center of Quantum Matter, Beijing 100190, China}

\begin{abstract}
We investigate the fine-grained uncertainty relations for qubit system by measurements
corresponding respectively to two and three spin operators.
Then we derive the general bound for a combination of two probabilities of projective measurements in mutually unbiased bases
in $d$-dimensional Hilbert space.
All of those uncertainty inequalities can be applied to construct different thermodynamic
cycles such that the violation of those inequalities will lead to
the violation of the second law of thermodynamics.
This reveals the relationship between fine-grained uncertainty and the second
law of thermodynamics.
\end{abstract}

\maketitle

\section{Introduction}
At the heart of quantum mechanics lies Heisenberg uncertainty principle
\cite{heisenberg}, which bounds the uncertainties about the outcomes of two incompatible measurements.
For example, it is well-known that for quantum mechanics,
if the momentum of a particle is predicted with certainty, when measuring its position,
all outcomes would occur which corresponds to
complete uncertainty. Originally the uncertainty principle is expressed by the Heisenberg-Robertson
relation \cite{robertson}:
\begin{equation}\label{1}
   \bigtriangleup R \cdot\bigtriangleup S\geq \frac{1}{2} |\langle[R,S]\rangle|,
\end{equation}
with standard deviations $\bigtriangleup R$ and $\bigtriangleup S$ for observable $R$ and $S$. There exist limitations
using the standard deviation as a measure of uncertainty, since the bound on the
right-hand side of relation (\ref{1}) depends on the state. To overcome this problem, Deutsch
\cite{deutsch} proposed a relationship to quantify uncertainty in terms of Shannon entropy.
Soon afterwards, an improved entropic uncertainty relation was established by Kraus \cite{kraus}
and then proved by Maassen and Uiffink \cite{maassen} with form
\begin{eqnarray}
H(R)+H(S)\geq\log_{2}\frac{1}{c},
\end{eqnarray}
where $H(R)$ denotes the Shannon entropy of the probability distribution of the outcomes when the observable
$R$ is measured. The entropic uncertainty relation provides us a way to quantify the uncertainties
with more than two measurements independent of the form of state. The entropic uncertainty relation
in the presence of quantum memory has also been presented in Ref.~\cite{berta}. So far many studies
on the entropic uncertainty relation have been done \cite{ruiz,ghirardi,wu,survey}.

Even though the entropic function is better to describe uncertainty than the standard deviation,
it is still a rather coarse way of measuring the uncertainty of a set of measurements. Entropy
is just a function of the probability distribution when a measurement is taken, and it cannot distinguish
the uncertainty inherent in obtaining any combination of outcomes for different measurements.
A new form of uncertainty relation, i.e., fine-grained uncertainty relation, was proposed by
Oppenheim and Wehner \cite{finegrained} to overcome this defect. For a set of measurements,
there exists a set of inequalities, one for each combination of possible outcomes:
\begin{equation}\label{2}
   \left\{\sum_{t=1}^{n}p(t)p(x^{(t)}|\rho)\leq\zeta_{\bm{x}}\Big|\bm{x}\in\mathbf{B}^{\times n} \right\},
\end{equation}
where $p(t)$ is the probability of choosing measurement labeled $t$, $p(x^{(t)}|\rho)$ is the
probability that we obtain the outcome $x^{(t)}$ when performing measurement $t$ on the state
$\rho$, the number of measurements is $n$,
$\bm{x}=(x^{(1)},\ldots,x^{(n)})$ is a combination of possible outcomes and
$\mathbf{B}^{\times n}$ is the set involving all possible combinations of outcomes.
Here, 
$\zeta_{\bm{x}}=\max_{\rho}\sum_{t=1}^n p(t)p(x^{(t)}|\rho)$,
 and the maximum is taken over all states allowed on a particular system.
 When $\zeta_{\bm{x}}<1$, we cannot obtain outcomes with certainty for all
measurements simultaneously.  A state that saturates the
inequality (\ref{2}) is called a ``maximally certain state".
Till now the fine-grained uncertainty relation has been applied in various aspects \cite{nonlocal,fineentropy}.
Remarkably, it is shown that the violation of the
fine-grained uncertainty implies the violation of the second law of thermodynamics \cite{violation}.

The explicit fine-grained uncertainty relation of a qubit is obtained with two Pauli operators \cite{finegrained}.
It is still necessary to study the general higher dimensional systems for two projective measurements
and the case with all three Pauli operators for a qubit as projective measurements.
In particular, it is not clear whether similar relationship between uncertainty relation and
the second law of thermodynamics in Ref.~\cite{violation} still exists generally.
In this paper we will study those problems.

This paper is organized as follows.
In section \uppercase\expandafter{\romannumeral2},
we investigate two-dimensional fine-grained uncertainty relations including two arbitrary spin operators and
provide a reasonable interpretation that the measurements $\{\sigma_x$, $\sigma_z\}$
behave as the best measurement basis for uncertainty relation.
In section \uppercase\expandafter{\romannumeral3},
we derive a general fine-grained uncertainty relation for mutually unbiased bases in $d$ dimension.
In section \uppercase\expandafter{\romannumeral4},
we propose the thermodynamic cycles and show that the violation of
generalized fine-grained uncertainty relations can also imply the violation of the second law of thermodynamics.
In section \uppercase\expandafter{\romannumeral5}, we have discussion and conclusion.

\section{Two-dimensional fine-grained uncertainty relations}
In two-dimensional Hilbert space an arbitrary pure state can be expressed as
$|\psi\rangle=\cos\frac{\theta}{2}|0\rangle+e^{i\phi}\sin\frac{\theta}{2}|1\rangle$ with $\theta\in[0,\pi]$ and $\phi\in[0,2\pi)$.
Here, we label the eigenbases of $\sigma_x$, $\sigma_y$ and $\sigma_z$
as $\{|+\rangle, |-\rangle\}$, $\{|\widetilde{+}\rangle, |\widetilde{-}\rangle\}$ and $\{|0\rangle, |1\rangle\}$.
We know that these four states, $\{|0\rangle,|1\rangle\}$ and $\{|+\rangle,|-\rangle\}$ŁŹ
are the well known BB84 states used for quantum key distribution \cite{bb}
which is unconditional secure because of the quantum uncertainty principle.
We propose that those two pairs of bases may enable the uncertainty to be optimal on average.
We can explain the uncertainty as follows: For an unknown state,
we expect to measure it accurately with two
measurements as far as possible. That is to say, we try to obtain the maximal bound for fine-grained
uncertainty relation. Taking a qubit as an example, due to the symmetry of the Bloch sphere, we just
consider the $x$-$z$ semi-plane, namely $\theta\in[0,\pi]$, $\phi=0$. The states
$|\psi\rangle=\cos\frac{\theta}{2}|0\rangle+\sin\frac{\theta}{2}|1\rangle$ with different $\theta$
appear with equal probability. Without loss of generality, we can first fix a measurement $\sigma_z$
and then pick an arbitrary operator $\sigma_n$, of which the angle between the two is $\alpha$ with
$\alpha\in[0,\pi]$. Thus the corresponding measurement probabilities are,
$p_{z\uparrow}=\cos^2\frac{\theta}{2}$, $p_{n\uparrow}=\cos^2\frac{\theta-\alpha}{2}$.
In order to take all states into account, we will compute the following integral
\begin{equation}\label{5}
  \int_0^{\pi}\frac{p_{z\uparrow}+p_{n\uparrow}}{\pi}d\theta=1+\frac{\sin\alpha}{\pi},
\end{equation}
where the maximum is reached when $\alpha=\pi/2$, i.e., $\sigma_n=\sigma_x$.
This gives an interpretation
that $\sigma_x$ and $\sigma_z$ denote the optimal measurements for the uncertainty relation.

It is well-reasoned to study the fine-grained uncertainty for $\sigma_x$ and $\sigma_z$.
Assume that these two Pauli operators are chosen with equal probability $1/2$, and for all pure states
$\rho=|\psi\rangle\langle\psi|$, the measurements probabilities satisfy the relation \cite{finegrained},
\begin{equation}\label{4}
  \frac{1}{2}p(0^{(x)}|\rho)+\frac{1}{2}p(0^{(z)}|\rho)\leq\frac{1}{2}+\frac{1}{2\sqrt{2}},
\end{equation}
with $0^{(x)}=|+\rangle$ and $0^{(z)}=|0\rangle$.
The equality of (\ref{4}) is satisfied when $\theta=\frac{\pi}{4}$ and $\phi=0$
corresponding to state $|\psi_m\rangle=\cos\frac{\pi}{8}|0\rangle+\sin\frac{\pi}{8}|1\rangle$
which is just an eigenstate of $(\sigma_x+\sigma_z)/\sqrt{2}$.
The two-dimensional Hilbert space exhibits remarkable geometrical properties: a pure state
corresponds to a point on the Bloch sphere \cite{book}. Thus $|\psi_m\rangle$ is located on the angle bisector between $x$-axis and $z$-axis. For other pairs
of outcomes $(0^{(x)},1^{(z)})$, $(1^{(x)},0^{(z)})$ and $(1^{(x)},1^{(z)})$, we can obtain the same
bound as inequality (\ref{4}).

Let us then consider the two-dimensional fine-grained uncertainty relation from another perspective.
Here we study the case including two arbitrary spin operators.
We choose two spin operators, $A=\bm{\sigma}\cdot\bm{m}$ and $B=\bm{\sigma}\cdot\bm{n}$,
in which $\bm{m}$ and $\bm{n}$ are unit vectors \cite{ghirardi}.
A normalized state $|\psi\rangle$ can be viewed as the eigenvector of $\bm{\sigma}\cdot\bm{k}$ projective to eigenvalue $+1$,
which means that $|\psi\rangle$ can be expressed as a unit vector $\bm{k}$ in the three-dimensional Euclidean space
(all these vectors form the so-called Bloch sphere). The corresponding measurement probabilities are
$p(\bm{m}_{\uparrow})=|\langle\bm{m}_{\uparrow}|\bm{k}_{\uparrow}\rangle|^2
=\frac{1}{2}(1+\bm{m}\cdot\bm{k})$ and $p(\bm{n}_{\uparrow})=\frac{1}{2}(1+\bm{n}\cdot\bm{k})$.
Thus $p(\bm{m}_{\uparrow})+p(\bm{n}_{\uparrow})=
1+\frac{1}{2}(\bm{m}+\bm{n})\cdot\bm{k} \leq \zeta$, in which $\zeta$
takes the maximum of $1+\frac{1}{2}(\bm{m}+\bm{n})\cdot\bm{k}$
over all vectors $\bm{k}$ forming the Bloch sphere.
When $\bm{k}$ is parallel to $\bm{m}+\bm{n}$, namely, when $\bm{k}$
lies in the direction of angle bisector between $\bm{m}$ and $\bm{n}$, the inequality takes the maximum.
$\zeta=1+\frac{1}{2}|\bm{m}+\bm{n}||\bm{k}|=1+\frac{|\bm{m}+\bm{n}|}{2}$. Then
\begin{equation}\label{12}
p(\bm{m}_{\uparrow})+p(\bm{n}_{\uparrow})\leq 1+\cos\frac{\gamma}{2},
\end{equation}
in which $\gamma\in(0,\pi)$ is the angle between $\bm{m}$ and $\bm{n}$.
If $\bm{m}\rightarrow \bm{x}$ and $\bm{n}\rightarrow \bm{z}$,
we find $|\bm{m}+\bm{n}|=\sqrt{2}$, with $\gamma={\pi}/{2}$. So we will reach the relation (\ref{4}).
From relation (\ref{12}), we can see that as the angle $\gamma$ becomes larger, the
bound for the fine-grained uncertainty relation becomes smaller.

\section{Fine-grained uncertainty relation for mutually unbiased bases}
In order to investigate the fine-grained uncertainty relation in higher dimensional Hilbert space,
we restrict our discussion to the projective measurements with mutually unbiased bases (MUBs) \cite{ivonovic}.
For dimension $d$, there can be at most $d+1$ MUBs which can be referred to as MUB-set.
It is shown that \cite{mub} the explicit construction of MUBs exists for prime power dimensions,
so we confine our study to a Hilbert space of prime power dimension.
The characterization of MUBs is that the squared overlaps between a state in one base of MUB-set and all states in the other bases are identical.
Considering the measurement in MUB, in case we detect a particular state in one base of MUB-set,
all outcomes of the measurement with another base will occur with equal probabilities.
Therefore, the MUBs are optimal bases to detect quantum uncertainty and thus can be used to quantum cryptography protocol \cite{xiong}.
In particular, the entropic uncertainty relation with MUBs has been explored in detail \cite{wu},
and the fine-grained uncertainty relations with MUBs are to be derived.
The simplest example of MUBs consists of  three Pauli matrices  in the case of spin-$1/2$ particle.
For general case,
let $\{|j\rangle\}_{j=0}^{d-1}$ and $\{|j^{(k)}\rangle\} (k=0,1,\ldots,d-1)$ denote a complete set of MUBs \cite{measurement}.
We use $k=\ddot{0}$ to label the base $\{|j\rangle\}$,
which are the eigenvectors of generalized Pauli operator $Z$, $Z|j\rangle=w^j|j\rangle, w=e^{2\pi i/d}$, and
\begin{equation}\label{6}
  |j^{(k)}\rangle=\frac{1}{\sqrt{d}}\sum_{l=0}^{d-1}w^{kl^2-2jl}|l\rangle.
\end{equation}
Next we will discuss how to obtain a fine-grained uncertainty relation in dimension $3$ with MUBs as an explicit example,
and then generalize to $d$-dimensional Hilbert space.

A three-dimensional pure state can be written as
\begin{eqnarray}
 |\Psi\rangle &=&\cos x_0|0\rangle+e^{i\varphi_1}\sin x_0\cos x_1|1\rangle
\nonumber \\
&& +e^{i\varphi_2}\sin x_0\sin x_1|2\rangle
\end{eqnarray}
with $x_0, x_1\in[0,\frac{\pi}{2}]$ and $\varphi_1, \varphi_2\in[0,2\pi)$.
Let $p(j^{(k)}|\rho)$ denote the probability of obtaining the outcome $j^{(k)}$ when taking the measurement with MUB
labeled by $k$ on pure state $\rho=|\Psi\rangle\langle\Psi|$. Then it is easily to obtain,
\begin{equation}\label{6}
  \frac{1}{2} p(0^{(\ddot{0})}|\rho)+\frac{1}{2}p(0^{(0)}|\rho)\leq \frac{1}{2}+\frac{1}{2\sqrt{3}}.
\end{equation}
The equality can be saturated when $\varphi_1=\varphi_2=0$, $x_1=\pi/4$, and $x_0=\frac{\pi}{4}-\frac{1}{2}\arcsin\frac{1}{\sqrt{3}}$.
When the inequality is saturated, the corresponding state has maximal certainty and
satisfies the condition $p(0^{(\ddot{0})}|\rho)=p(0^{(0)}|\rho)$, which means
that the angle between the state and base $|0\rangle$ is the same as the one between the state and base $|0^{(0)}\rangle$.
Similarly, any combination of two outcomes from different measurements has the equal upper bound as presented above.

Now, let us see the fine-grained uncertainty relation for MUBs 
in $d$-dimensional Hilbert space. We can obtain the uncertainty relation as,
\begin{equation}\label{7}
   \frac{1}{2}p(0^{(\ddot{0})}|\rho)+ \frac{1}{2}p(0^{(0)}|\rho)\leq \frac{1}{2}+\frac{1}{2\sqrt{d}}.
\end{equation}
This can be proven in the similar way as the three-dimensional case.
A pure state can be described generally as
$|\Psi\rangle=\cos x_0|0\rangle+e^{i\varphi_1}\sin x_0\cos x_1|1\rangle+\cdots
+e^{i\varphi_{d-1}}\sin x_0\sin x_1\cdots\sin x_{d-2}|d-1\rangle$.
The probabilities $p(0^{(\ddot{0})}|\rho)$ and $p(0^{(0)}|\rho)$ are easily calculated.
The inequality $\sqrt{n}\sin x+\cos x \leq\sqrt{n+1}$ is used to determine the final bound.
The same upper bound as relation (\ref{7}) holds for any other pairs of outcomes from different measurements.

\section{Relationship with the second law of thermodynamics}
\subsection{The case of three-dimension}
The measurement operators composed of the MUBs are noncommuting.
It is significant to establish fine-grained uncertainty relation for mutually unbiased bases.
Furthermore, the uncertainty relation (\ref{6}) can also be applied to construct a thermodynamic cycle
similar as that in \cite{violation,vedral}.

\begin{figure}[tp]
 \centering
  \includegraphics[width=0.40\textwidth, height=6 cm]{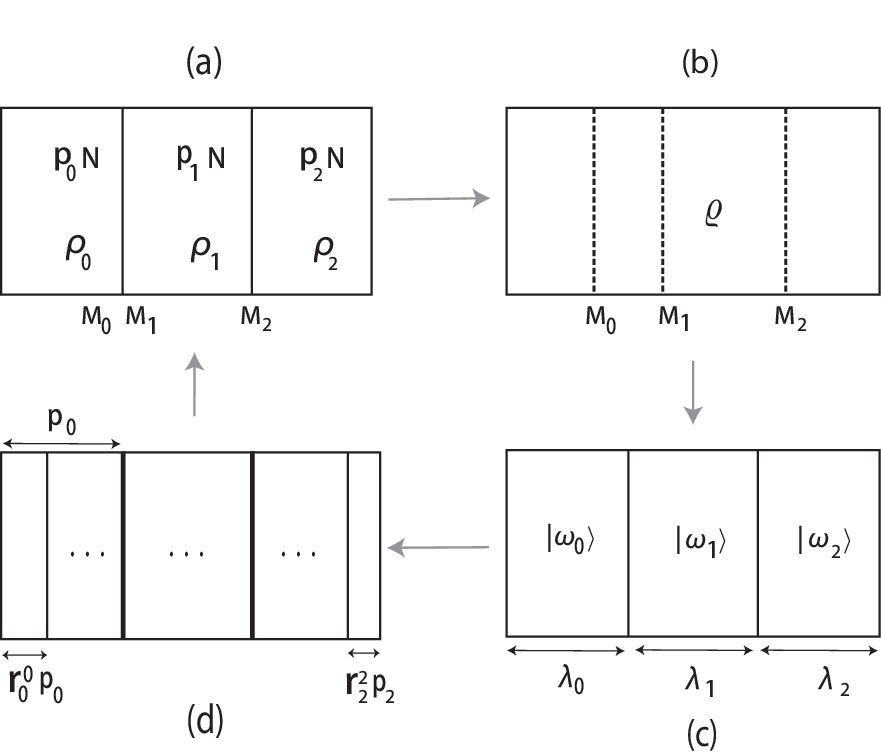}\\
\caption{A thermodynamic cycle. There are two paths from the initial state to the completely mixed state. The first path (a)$\rightarrow$(b)
denotes the first part of the cycle. This process is irreversible. The second path (a)$\rightarrow$(d)$\rightarrow$(c)$\rightarrow$(b) is a reversible process
 and its reversed process denotes the second part of the cycle.}\label{f1}
\end{figure}

As shown in FIG.~\ref{f1}, at the very beginning, we prepare three kinds of particles $\rho_0$, $\rho_1$ and $\rho_2$ with the numbers of $p_0N$, $p_1N$ and $p_2N$, respectively,
\begin{eqnarray}
\rho_0&=&\frac{|0\rangle\langle 0|+|0^{(0)}\rangle\langle 0^{(0)}|}{2},\
\rho_1=\frac{|1\rangle\langle 1|+|1^{(0)}\rangle\langle 1^{(0)}|}{2},\nonumber\\
\rho_2&=&\frac{|2\rangle\langle 2|+|2^{(0)}\rangle\langle 2^{(0)}|}{2}.
\end{eqnarray}
They are put into a vessel divided into three volumes with $p_0V$, $p_1V$ and $p_2V$ by partitions.
In this situation, we use semi-transparent membranes that were imagined by von Neumann \cite{von} and Peres \cite{peres}.
A membrane labeled by $M_0$ is opaque to a normalized state $|e_0\rangle$ and transparent to the
other orthogonal normalized states $|e_1\rangle$ and $|e_2\rangle$.
The process that a state passes through the membrane corresponds to a projective measurement in base $e=\{|e_0\rangle, |e_1\rangle, |e_2\rangle\}$.
After we replace the left partition by two membranes $M_0$ and $M_1$ and the right one by $M_2$,
the state $\rho_0$ in the left side is reflected by $M_0$ to become $|e_0\rangle$ with probability $p(e_0|\rho_0)$
and passes through with probability $p(e_1|\rho_0)+p(e_2|\rho_0)=1-p(e_0|\rho_0)$.
Similarly, the state $\rho_0$ is projected onto $|e_1\rangle$ with probability $p(e_1|\rho_0)$ and projected onto $|e_2\rangle$ with probability $p(e_2|\rho_0)$. Here, we define
\begin{eqnarray}
p(e_j|\rho)\equiv\langle e_j|\rho|e_j\rangle \quad j=0,1,2.
\end{eqnarray}
The states $\rho_1$ and $\rho_2$ experience the similar process.
Thus, these inserted membranes will move apart until they are in equilibrium.
A membrane $M_j$ will not move until the density of the particles in state $|e_j\rangle$ on both sides are equal.
If we let $p_0=p_1=p_2=1/3$, the state in the vessel can be written as $\varrho\equiv\frac{1}{3}(\rho_0+\rho_1+\rho_2)=\mathbf{1}/3$ when the equilibrium is reached.
Because the state $\varrho$ is the complete mixed state, we can write $p(e_j|\varrho)=\frac{1}{3}$ for $j=0,1,2.$
The membranes moving process enables us to extract work $W_1$ from the system.  The extracted work can be calculated as \cite{vedral},
\begin{eqnarray}
&&W_1 = -\sum_{i=0}^2 p_i\log p_i-\sum_{j=0}^2 p(e_j|\varrho)\log p(e_j|\varrho)\nonumber\\
&+&[p_1 p(e_0|\rho_1)+p_2 p(e_0|\rho_2)]\log[p_1 p(e_0|\rho_1)+p_2 p(e_0|\rho_2)]\nonumber\\
&+&[p_1 p(e_1|\rho_1)+p_2 p(e_1|\rho_2)]\log[p_1 p(e_1|\rho_1)+p_2 p(e_1|\rho_2)] \nonumber\\
&+&[p_0 p(e_2|\rho_0)+p_1 p(e_2|\rho_1)]\log[p_0 p(e_2|\rho_0)+p_1 p(e_2|\rho_1)] \nonumber\\
&+& p_0 p(e_0|\rho_0)\log[p_0 p(e_0|\rho_0)] + p_0 p(e_1|\rho_0)\log[p_0 p(e_1|\rho_0)]\nonumber\\
&+& p_2 p(e_2|\rho_2)\log[p_2 p(e_2|\rho_2)].
\end{eqnarray}
In the expression of work, we omit the physical factor $NkT\ln2$.

Next we describe a path from the mixed state $\varrho$ to the initial state to constitute a closed thermodynamic cycle.
The membranes $M_0, M_1, M_2$ are then taken away. According the the eigen-decomposition $\varrho=\sum_j \lambda_j|\omega_j\rangle\langle \omega_j|$,
New membranes that can distinguish its eigenvectors are inserted to separate the state into $|\omega_j\rangle$, each with volume $\lambda_jV$. This process needs work \cite{vedral,von}.
According to the decompositions $\rho_0=\sum_j r_j^0|\chi_j^0\rangle\langle\chi_j^0|$,
$\rho_1=\sum_j r_j^1|\chi_j^1\rangle\langle\chi_j^1|$ and $\rho_2=\sum_j r_j^2|\chi_j^2\rangle\langle\chi_j^2|$,
we insert partitions to subdivide the separated volumes into smaller sizes proportional to weights $r_j^0$, $r_j^1$ and $r_j^2$.
The conversion from the pure state $|\omega_j\rangle$ to $|\chi_j^0\rangle$, $|\chi_j^1\rangle$ or $|\chi_j^2\rangle$ needs no work.
Finally we retrieve $\rho_0$, $\rho_1$ and $\rho_2$ by mixing their respective components.
This retrieval process needs work \cite{violation,vedral}
\begin{equation}\label{10}
   W_2=S(\varrho)-\sum_{j=0}^2 p_j S(\rho_j),
\end{equation}
where $S(\varrho)$ denotes the von Neumann entropy. Thus, after this thermodynamic loop
the total work is given by
\begin{eqnarray}
W_{net} &=&W_1-W_2 =H_{b}\left(\frac{1}{2}+\frac{1}{2\sqrt{3}}\right)\nonumber\\
  & -&{\frac{1}{3} H_{b}\left[\frac{1}{2}p(0^{(\ddot{0})}|\rho_{e_0})+\frac{1}{2}p(0^{(0)}|\rho_{e_0})\right]}\nonumber\\
   &-&{\frac{1}{3} H_{b}\left[\frac{1}{2}p(0^{(\ddot{0})}|\rho_{e_1})+\frac{1}{2}p(0^{(0)}|\rho_{e_1})\right]} \nonumber\\
   &-&{\frac{1}{3} H_{b}\left[\frac{1}{2}p(2^{(\ddot{0})}|\rho_{e_2})+\frac{1}{2}p(2^{(0)}|\rho_{e_2})\right] }
\end{eqnarray}
where $H_{b}(p)=-p\log p-(1-p)\log(1-p)$ is the binary entropy and $\rho_{e_{j}}=|e_{j}\rangle\langle e_{j}|$ for $j=0,1,2$.

The monotonicity of $H_{b}(p)$ and relation (\ref{6}) imply $W_{net}\leq0$.
If the uncertainty relation is violated, namely, the quantity in right hand side of inequality (\ref{6})
becomes higher, then $W_{net}>0$ may occur. That means net work can
be extracted from a thermodynamic cycle and thus the second law of thermodynamics is violated.
The uncertainty relations are directly inferred from the mathematical formalism of quantum theory,
and this microscopic inequality seems not to relate to the macroscopic theory of thermodynamics on the surface.
However, our result gives a thermodynamic operational meaning for uncertainty principle.
In FIG.~\ref{f1}, the first part of cycle from (a) to (b) involves irreversible projective measurement,
in which the uncertainty relations are inserted into the work expression,
while the returning process (c)$\rightarrow$(d)$\rightarrow$(a) is reversible, where the work does not contain the uncertainty relation.
Hence, the uncertainty principle is related to the reversibility, i.e., the direction of thermodynamic process,
which is the core of the second law of thermodynamics.
The above hypothetical thermodynamic cycle with membranes plays an important role in connecting quantum theory with thermodynamics.
The violation of quantum superposition principle, quantum state discrimination and linear evolution of quantum states
can also imply the violation of the second law, which has been studied in \cite{colloquium}.

\subsection{The case of d-dimension}
\begin{figure}[tp]
 \centering
  \includegraphics[width=0.30\textwidth, height=4 cm]{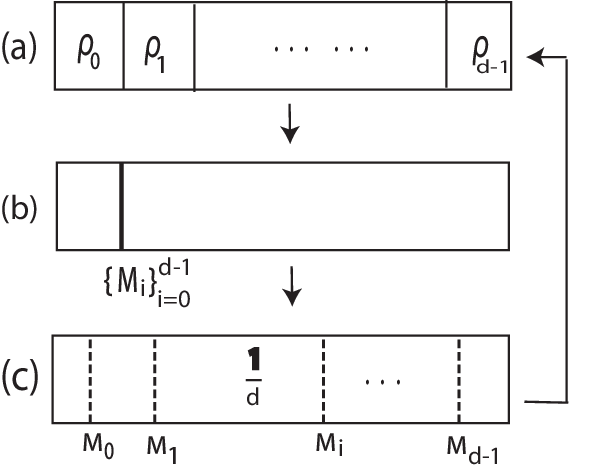}\\
\caption{The thermodynamic cycle for d-dimension.}\label{f2}
\end{figure}

Moreover, the similar cycle can be generally extended to $d$-dimensional situation.
At this point, the preparation is shown in Fig.~\ref{f2} (a): the states
$\rho_j=\frac{1}{2}(|j\rangle\langle j|+|j^{(0)}\rangle\langle j^{(0)}|), (j=0,\ldots,d-1)$ are separated by $d-1$ partitions.
The first partition can be replaced by a set of $d$ membranes and the others are directly removed.
The membranes then move due to the penetration of particles.
In equilibrium the state in the volume  becomes the complete mixed state $\varrho=\mathbf{1}/d$, as shown in Fig.~\ref{f2} (c).
The reversed process from (c) to (a) is similarly designed as in Fig.~\ref{f1}:
$\varrho\rightarrow\{\lambda_i, |\omega_i\rangle\}\rightarrow\{p_j r_k^j, |\chi_k^j\rangle\}\rightarrow\{p_j, \rho_j\}$,
where $\{r_k^j, |\chi_k^j\rangle\}$ and $\{\lambda_i, |\omega_i\rangle\}$ are the eigenvalue decomposition of $\rho_j$ and $\varrho$ \cite{vedral}.
After the whole cycle, we can obtain
\begin{eqnarray}
&& W_{net}=H_b \left(\frac{1}{2}+\frac{1}{2\sqrt{d}}\right)\nonumber\\
   &-&{\frac{1}{d} H_b \left[\frac{1}{2}p(0^{(\ddot{0})}|\rho_{e_0})+\frac{1}{2}p(0^{(0)}|\rho_{e_0})\right]} -\cdots\nonumber\\
   &-&{\frac{1}{d} H_b\left[\frac{1}{2}p(0^{(\ddot{0})}|\rho_{e_{d-1}})+\frac{1}{2}p(0^{(0)}|\rho_{e_{d-1}})\right] }.
\end{eqnarray}
If the d-dimensional uncertainty relation (\ref{7}) is violated,
then the terms in square brackets in the above relation may exceed the bound $\frac{1}{2}+\frac{1}{2\sqrt{d}}$.
Thus it is possible to get $W_{net}>0$, leading to the violation of the second law of thermodynamics.

\subsection{The case with adjustable parameter}
By choosing arbitrary two incompatible measurements, we are able to construct corresponding uncertainty relations.
In qubit case, we can consider the uncertainty relation with adjustable parameter,
i.e., $\sigma _z$ and operator $\sigma _n$ with a tunable direction are chosen such that the bound of the uncertainty relation can be adjustable.
In this case, the violation of the uncertainty relation with adjustable parameter always leads to the violation of the second law of thermodynamics.
We have already obtained a fine-grained relation (\ref{12}) for any two spin operators $A=\bm{\sigma}\cdot\bm{m}$ and $B=\bm{\sigma}\cdot\bm{n}$.
Now we consider two measurements $\sigma_z$ and $\sigma_n$, in which the direction of $\sigma_n$ is
$\bm{n}=\{\sin\theta\cos\phi, \sin\theta\sin\phi, \cos\theta\}$.
From Eq. (\ref{12}), for pure state $\rho=|\psi\rangle\langle\psi|$, we have,
\begin{equation}\label{a}
  \frac{1}{2}p(0^z|\rho)+\frac{1}{2}p(0^n|\rho)\leq \frac{1}{2}(1+\cos\frac{\theta}{2}),
\end{equation}
\begin{equation}\label{b}
  \frac{1}{2}p(1^z|\rho)+\frac{1}{2}p(1^n|\rho)\leq \frac{1}{2}(1+\cos\frac{\theta}{2}),
\end{equation}
where $0^n= \cos\frac{\theta}{2}e^{-i\phi/2}|0\rangle+\sin\frac{\theta}{2}e^{i\phi/2}|1\rangle$
and $1^n= -\sin\frac{\theta}{2}e^{-i\phi/2}|0\rangle+\cos\frac{\theta}{2}e^{i\phi/2}|1\rangle$.

We can apply the relations (\ref{a}) and (\ref{b}) into the thermodynamic cycle with the preparation,
\begin{eqnarray*}
&&\rho_0=\frac{|0\rangle\langle 0|+|0^{n}\rangle\langle 0^{n}|}{2},\qquad
  \rho_1=\frac{|1\rangle\langle 1|+|1^{n}\rangle\langle 1^{n}|}{2},\\
&&\varrho=\frac{\rho_0+\rho_1}{2}=\frac{\mathbf{1}}{2}.
\end{eqnarray*}
After a similar cycle as stated above, the total work can be calculated as,
\begin{eqnarray}
 W_{net} &=&
   H\left(\frac{1}{2}(1+\cos\frac{\theta}{2})\right)\nonumber\\
  && -\frac{1}{2} H\left[\frac{1}{2}p(0^z|\rho_{e_1})+\frac{1}{2}p(0^n|\rho_{e_1})\right]\nonumber\\
   &&-\frac{1}{2} H\left[\frac{1}{2}p(1^z|\rho_{e_2})+\frac{1}{2}p(1^n|\rho_{e_2})\right].
\end{eqnarray}
Therefore, in 2-dimensional Hilbert space, we can always construct the relationship between the fine-grained uncertainty and the second law of thermodynamics.
That is to say, the violation of the fine-grained uncertainty relations with two arbitrary spin operators leads to
the violation of the second law.
In Ref.~\cite{violation}, the general result in dimension two is given by
\begin{equation}\label{w}
 W_{net} =\sum_i p_i S(\rho_i)-S(\varrho)+1-\frac{1}{2}H(\zeta_{(f_0,g_0)})-\frac{1}{2}H(\zeta_{(f_1,g_1)}),
\end{equation}
in which $\zeta_{(f_i,g_i)} (i=0,1)$ is the fine-grained uncertainty bound for measurement operators $f_i$ and $g_i$.
However, from this result we cannot obviously conclude the violation of uncertainty will lead to
the violation of the second law of thermodynamics.
In this paper, we derive a general and obvious relation (19) for any spin operators in qubit case.

\subsection{Extension to more than two probabilities}
In the above discussion, we just consider the uncertainty relations with two measurements.
It is also interesting to investigate a set of arbitrary number of measurements as those for quantum key distributions \cite{six}.
As a simple example, we consider a combination of three outcomes in two-dimensional Hilbert space, namely,
we choose measurements $\sigma_x$, $\sigma_y$ and $\sigma_z$ with equal probability $1/3$, then one obtains
\begin{eqnarray}
\frac{1}{3}[p(0^{(x)}|\rho)+p(0^{(y)}|\rho)+p(0^{(z)}|\rho)]  \leq \frac{1}{2}+\frac{1}{2\sqrt{3}},\label{17}
\end{eqnarray}
where $\rho=|\psi\rangle\langle\psi|$ is a pure state and the equality is saturated when 
$\theta=\arcsin\sqrt{{2}/{3}}$, $\phi={\pi}/{4}$.
Therefore, in the Bloch sphere representation the maximally certain state lies
on the body diagonal formed by $x$-axis, $y$-axis and $z$-axis.
For other combinations of outcomes, we can obtain the same upper bound.
Similarly this inequality can also be applied to the above mentioned thermodynamic cycle.
We can prepare the initial state $\rho_0=\frac{1}{3}(|+\rangle\langle+|+|\widetilde{+}\rangle\langle\widetilde{+}|+|0\rangle\langle0|)$
and $\rho_1=\frac{1}{3}(|-\rangle\langle-|+|\widetilde{-}\rangle\langle\widetilde{-}|+|1\rangle\langle1|)$ with one partition.
After substituting two membranes for the partition and proceeding the whole cycle, the net work gives
\begin{eqnarray}
W_{net} &=&H_{b}\left(\frac{1}{2}+\frac{1}{2\sqrt{3}}\right)\nonumber\\
  & -&{\frac{1}{2} H_{b}\left[\frac{1}{3}[p(0^{(x)}|\rho)+p(0^{(y)}|\rho)+p(0^{(z)}|\rho)]\right]}\nonumber\\
   &-&{\frac{1}{2} H_{b}\left[\frac{1}{3}[p(1^{(x)}|\rho)+p(1^{(y)}|\rho)+p(1^{(z)}|\rho)]\right]}.
   \nonumber \\
\end{eqnarray}
Thus if we are able to go beyond the uncertainty relations, the operational consequences cause the violation of the
second law of thermodynamics.

\section{Discussion and Conclusion}
Uncertainty principle is a fundamental property of quantum mechanics.
Furthermore, it has been continuously studied and represented in various forms.
It is interesting to find that the uncertainty principle can be related with the second law of thermodynamics.
This is demonstrated by constructing a thermodynamic circle such that the violation of the uncertainty relation may lead to the violation of the second law.
The reason behind this relationship is that although the uncertainty relation is derived from the mathematical formalism of quantum theory,
it holds a macroscopical physical implication by means of a thermodynamic cycle.
The compatibility between the uncertainty principle
of quantum mechanics and the second law of thermodynamics of statistical mechanics may
create some new restrictions acting as new fundamental principle of physics.
However, even we have shown that those relationship may exist for the fine-grained uncertainty relations,
it is still unclear how to construct various thermodynamic circles to involve other forms of uncertainty relations into the total thermodynamic work.

In summary we derive the fine-grained uncertainty relations for mutually unbiased bases.
In a qubit system, the fine-grained uncertainty relation for any two spin operators is obtained.
Thanks to the geometrical property of qubit, we reduce the problem to geometrical one,
using the correspondence between state vectors and points on the Bloch sphere.
Then we generalize the inequality to d-dimensional case where two projective measurements with MUBs are considered.
Since MUBs hold the special property that arbitrary two states from different bases have the same overlap,
we can obtain a general bound $\frac{1}{2}+\frac{1}{2\sqrt{d}}$ for any combination of two outcomes.
Furthermore, our new inequalities can be employed to a general thermodynamic cycle,
from which we derive an expression of total work which contains the quantum uncertainty.
Finally we discuss the fine-grained uncertainty relation for three Pauli operators.
However, the result for more than two measurements in higher-dimensional space is to be explored.

\begin{acknowledgments}
We thank Yu-Ran Zhang and Ling-An Wu for critical reading this paper and valuable discussions.
This work is supported by ``973" program (2010CB922904), NSFC (11175248), and
grants from Chinese Academy of Sciences.
\end{acknowledgments}



\begin{thebibliography}{99}
\bibitem{heisenberg} W. Heisenberg, Z. Phys. {\bf43}, 172 (1927).

\bibitem{robertson} H. P. Robertson, Phys. Rev. {\bf34}, 163 (1929).

\bibitem{deutsch} D. Deutsch, Phys. Rev. Lett. {\bf50}, 631 (1983).

\bibitem{kraus} K. Kraus, Phys. Rev. D {\bf35}, 3070 (1987).

\bibitem{maassen} H. Maassen, and J. B. M. Uffink, Phys. Rev. Lett. {\bf60}, 1103 (1988).

\bibitem{berta} M. Berta, M. Christandl, R. Colbeck, J. M. Renes, and R. Renner, Nat. Phys. {\bf6}, 659 (2010).

\bibitem{ruiz} J. Sanchez-Ruiz, Phys. Lett. A {\bf244}, 189 (1998).

\bibitem{ghirardi}  G. Ghirardi, L. Marinatto, and R. Romano, Phys. Lett. A {\bf317}, 32 (2003).

\bibitem{wu} S. J. Wu, S. X. Yu, and K. Molmer, Phys. Rev. A {\bf79}, 022104 (2009).

\bibitem{survey} S. Wehner, and A. Winter, New J. Phys. {\bf12}, 025009 (2010).

\bibitem{finegrained} J. Oppenheim, and S. Wehner, Science {\bf330}, 1072 (2010).

\bibitem{nonlocal} Ansuman Dey, T. Pramanik, and A. S. Majumdar, Phys. Rev. A {\bf87}, 012120 (2013).

\bibitem{fineentropy} T. Pramanik, P. Chowdhury, and A. S. Majumdar,  Phys. Rev. Lett. {\bf110}, 020402 (2013).

\bibitem{violation} E. Hanggi, and S. Wehner, Nature Commun. {\bf4}, 1670 (2013).

\bibitem{bb} C. H. Bennett, and G. Brassard, in Proceedings of IEEE International
Conference on Computers, Systems, and Signal Processing (IEEE, New York/Bangalore, 1984), pp. 175.

\bibitem{book} M. A. Nielsen, and I. L. Chuang, \emph{Quantum Computation and Quantum Information}
(Cambridge University Press, Cambridge, England, 2000).

\bibitem{ivonovic} I. D. Ivonovic, J. Phys. A {\bf14}, 3241 (1981).

\bibitem{mub} S. Bandyopadhyay, P. O. Boykin, V. Roychowdhury, and F. Vatan, Algorithmica {\bf34}, 512 (2002).

\bibitem{xiong}Z. X. Xiong, H. D. Shi, Y. N. Wang, L. Jing, J. Lei,
L. Z. Mu, and H. Fan,
Phys. Rev. A {\bf 85}, 012334 (2012).

\bibitem{wu}S. J. Wu, S. X. Yu, and K. Molmer, Phys. Rev. A {\bf 79}, 022104 (2009).

\bibitem{measurement} A. Kalev, A. Mann, and M. Revzen, Phys. Rev. Lett. {\bf110}, 260502 (2013).

\bibitem{vedral} K. Maruyama, C. Brukner, and V. Vedral, J. Phys. A Math. Gen. {\bf38}, 7175 (2005).

\bibitem{von} J. von Neumann, \emph{Mathematical Foundations of Quantum Mechanics} (Princeton University Press, 1955).

\bibitem{peres} A. Peres, \emph{Quantum Theory: Concepts and Methods. Fundamental Theories of Physics} (Kluwer Academic, 1993).

\bibitem{colloquium} K. Maruyama, F. Nori, and V. Vedral. Rev. Mod. Phys. {\bf81}, 1 (2009).

\bibitem{six} D. Bruss, Phys. Rev. Lett. {\bf81}, 3018 (1998).

\end{thebibliography}
\end{document}